\documentclass[reprint,aps,prb,amsmath,superscriptaddress,bibnotes,longbibliography]{revtex4-2}
\usepackage{mathptmx,newtxtext,newtxmath,xspace}
\usepackage{amsbsy,bm,bbold}
\usepackage{graphicx,color,xcolor}
\usepackage{fancyhdr}
\usepackage{comment}
\usepackage[colorlinks=true, 
linkcolor=blue, 
urlcolor=blue,
citecolor=blue]{hyperref}
\usepackage{psfrag}
\usepackage{enumitem}

\DeclareTextCompositeCommand{\r}{OT1}{A}{%
	\leavevmode\vbox{%
		\offinterlineskip
		\ialign{\hfil##\hfil\cr\char23\cr\noalign{\kern-1.15ex}A\cr}%
	}%
}
\usepackage[english]{babel}
\begin{document}

\title{Evolution of spin excitations in superconducting La$_{2-x}$Ca$_{x}$CuO$_{4-\delta}$ from the underdoped to the heavily overdoped regime}

\widetext
\date{\today}
	
\author{S. Hameed}
\affiliation{Max Planck Institute for Solid State Research, Heisenbergstrasse 1, 70569 Stuttgart, Germany}
\author{Y. Liu}
\affiliation{Max Planck Institute for Solid State Research, Heisenbergstrasse 1, 70569 Stuttgart, Germany}
\author{M. Knauft}
\affiliation{Max Planck Institute for Solid State Research, Heisenbergstrasse 1, 70569 Stuttgart, Germany}
\author{K. S. Rabinovich}
\affiliation{Max Planck Institute for Solid State Research, Heisenbergstrasse 1, 70569 Stuttgart, Germany}
\author{G. Kim}
\affiliation{Max Planck Institute for Solid State Research, Heisenbergstrasse 1, 70569 Stuttgart, Germany}
\author{G. Christiani}
\affiliation{Max Planck Institute for Solid State Research, Heisenbergstrasse 1, 70569 Stuttgart, Germany}
\author{G. Logvenov}
\affiliation{Max Planck Institute for Solid State Research, Heisenbergstrasse 1, 70569 Stuttgart, Germany}
\author{F. Yakhou-Harris}
\affiliation{European Synchrotron Radiation Facility, Boîte Postale 220, F-38043 Grenoble, France}
\author{A. V. Boris}
\affiliation{Max Planck Institute for Solid State Research, Heisenbergstrasse 1, 70569 Stuttgart, Germany}
\author{B. Keimer}
\affiliation{Max Planck Institute for Solid State Research, Heisenbergstrasse 1, 70569 Stuttgart, Germany}
\author{M. Minola}
\affiliation{Max Planck Institute for Solid State Research, Heisenbergstrasse 1, 70569 Stuttgart, Germany}

\begin{abstract}
We investigate high-energy spin excitations in hole-doped \(\mathrm{La}_{2-x}\mathrm{Ca}_x\mathrm{CuO}_{4-\delta}\) films across a broad Ca doping range (\(x = 0.05\)–\(0.50\)) using resonant inelastic x-ray scattering (RIXS). Polarization analysis and incident-photon energy detuning measurements confirm the persistence of collective paramagnon excitations up to \(x = 0.50\). Consistent with previous studies on other cuprate families, we observe a pronounced crossover near \(x = 0.15\), where paramagnon spectral weight is transferred to incoherent spin-flip excitations associated with the particle-hole continuum. The overall behavior of paramagnons in LCCO resembles that in other hole-doped cuprates and appears insensitive to the persistence of superconductivity at high doping levels in LCCO—up to at least \(x = 0.50\), as demonstrated in prior work. These findings support the view that high-energy magnetic excitations probed by RIXS are not a major contributor to superconducting pairing, in line with theories of spin-fluctuation mediated superconductivity.

\end{abstract}
\maketitle

\section{Introduction}
High-temperature superconducting cuprates remain one of the most intriguing systems in condensed matter physics, with the underlying mechanism of superconductivity still the subject of active debate \cite{Keimer2015}. These materials exhibit a rich phase diagram as a function of hole doping, encompassing antiferromagnetism, superconductivity, a pseudogap phase, charge order, and strange metal behavior. The interplay between these phases and their connection to superconductivity remains the focus of intense discussions and research efforts.

One of the striking features of cuprate superconductors is the dome-shaped dependence of the superconducting transition temperature \( T_c \) on doping. As hole concentration increases beyond optimal doping, \( T_c \) decreases and eventually vanishes in the heavily overdoped regime, giving rise to a Fermi liquid phase. Transport and thermodynamic data point to a quantum phase transition marking the end of the pseudogap phase at a hole doping level $p^* = 0.19$ \cite{Proust2019}, although this has been challenged by a recent report \cite{Nicholls2025}. This quantum critical point (QCP) is believed to underpin the non-Fermi-liquid behavior and the emergence of the ``strange metal'' phase characterized by linear-in-temperature resistivity. The critical fluctuations emanating from this point are believed to play a pivotal role in shaping the electronic properties of the cuprates, possibly even contributing to the pairing mechanism responsible for high-\( T_c \) superconductivity.

However, the precise nature of the QCP remains hotly debated. A central question is whether the criticality arises from the suppression of spin-related order, such as antiferromagnetism or spin-density waves, or from instabilities in the charge sector, such as charge-density waves (CDW) - or possibly both spin and charge. 
In the charge sector, recent work on the clean underdoped cuprate YBa$_2$Cu$_4$O$_8$ (Y124) revealed that CDW order onsets in a second-order-like fashion at the pseudogap transition temperature, suggesting a close connection between the two phenomena \cite{Betto2025}. Consistent with this finding, recent work on YBa$_2$Cu$_3$O$_{6+x}$ (YBCO) indicates robust CDW correlations for doping levels up to $p^*$ \cite{Zhou2025,ZhouA2025}. Furthermore, the intensity and characteristic energy of charge density fluctuations have been reported to exhibit a maximum and minimum respectively at $p^*$ \cite{Arpaia2023}.
On the other hand, in the spin sector, recent nuclear magnetic resonance studies on La$_{2-x}$Sr$_{x}$CuO$_4$ (LSCO) under high magnetic field and La$_{1.8-x}$Eu$_{0.2}$Sr$_{x}$CuO$_4$ (Eu-LSCO) in zero magnetic field have shown that the static spin-stripe order is closely tied to the pseudogap phase and ends at the pseudogap phase boundary \cite{Frachet2020,Missiaen2025}. However, resonant inelastic x-ray scattering (RIXS) studies on several cuprate families have shown that short-range spin excitations or paramagnons persist well into the overdoped regime \cite{LeTacon2011,Dean2013,LeTacon2013,Wakimoto2015,Minola2015,Monney2016,Meyers2017,Minola2017,Peng2018,Zhang2022}, with a strong crossover occurring near optimal doping where a loss of paramagnon spectral weight to incoherent spin-flip excitations of the particle-hole continuum is observed \cite{Minola2015,Minola2017}.

\begin{figure}
	\includegraphics[width=0.45\textwidth]{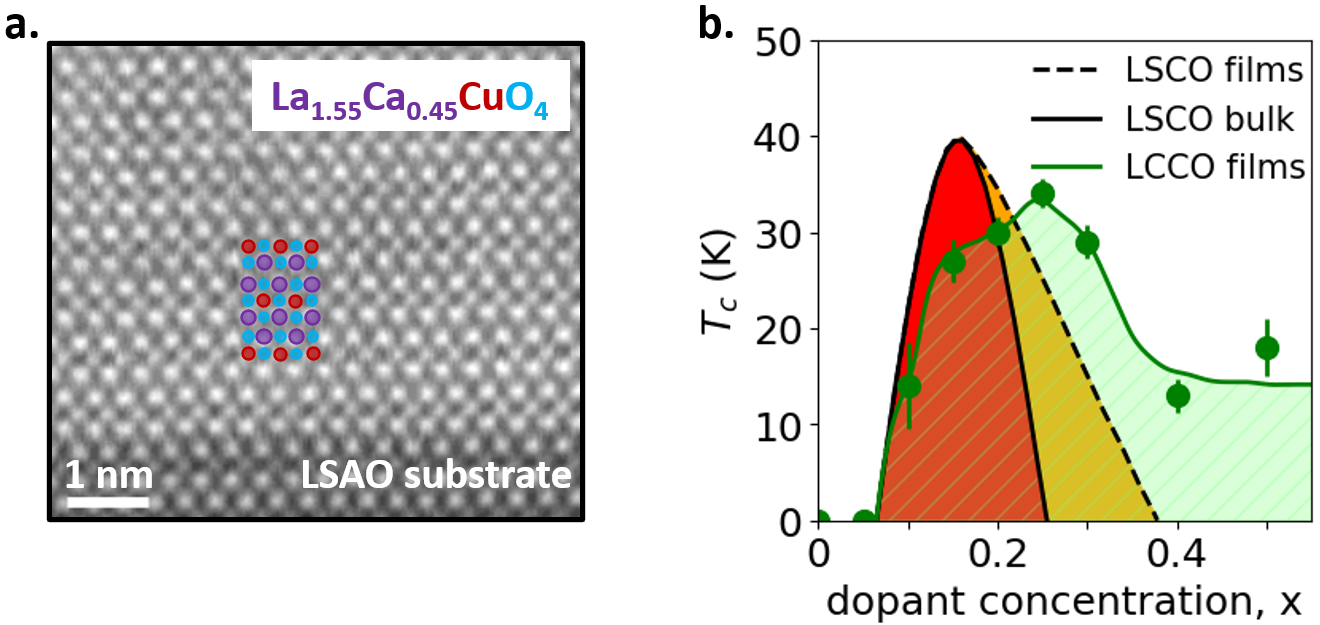}
	\caption{(a) Scanning transmission electron microscopy (STEM) image of an LCCO film with \( x = 0.45 \), demonstrating high crystallinity. (b) Superconducting \( T_c \) as a function of doping in 10-unit-cell-thick LCCO films, compared to bulk and thin-film LSCO \cite{Tallon1995,Dean2013,Sato2000,Kim2017}. Figures adapted from~\cite{Kim2021}.}
	\label{figPD}
\end{figure}

Recently, some of us succeeded in growing high-quality thin films of La$_{2-x}$Ca$_{x}$CuO$_{4-\delta}$ (LCCO) with Ca doping levels extending up to \(x = 0.50\)~\cite{Kim2021} [Fig.~\ref{figPD}(a)]. Likely due to the reduced structural disorder arising from the nearly identical ionic radii of La and Ca, superconductivity in this system persists up to at least \(x = 0.50\), with \(T_c \sim 15\)~K—well beyond what is typically observed in systems with higher disorder, such as LSCO [Fig.~\ref{figPD}(b)]. This extended superconducting phase diagram in LCCO provides an opportunity to test the sensitivity of spin and charge-related phase boundaries to the robustness of the superconducting state. In the charge sector, our recent resonant x-ray scattering and x-ray diffraction studies revealed that electronic CDW order exists in LCCO only up to a Ca doping of \(x \sim 0.20\), although a superstructure associated with oxygen-vacancy ordering persists beyond this doping level~\cite{Hameed2024}. In this paper, we systematically investigate the evolution of spin excitations in LCCO as a function of Ca doping using RIXS. Through polarization-resolved and incoming-photon-energy-dependent measurements, we demonstrate that dispersive spin excitations persist deep into the overdoped regime, up to at least \( x = 0.50 \). They start as Raman-like collective paramagnon excitations at low doping and progressively become more incoherent and fluorescent in nature at higher doping levels. In particular, a pronounced crossover near $x = 0.15$ is observed, characterized by a transfer of paramagnon spectral weight to incoherent spin-flip excitations associated with the particle-hole continuum, in agreement with observations in other cuprate families. Notably, the doping evolution closely mirrors that seen in other hole-doped cuprates and appears largely unaffected by the extended superconducting dome in LCCO [Fig.~\ref{figPD}(b)]. These results support the view that high-energy magnetic excitations probed by RIXS are marginal to the superconducting pairing mechanism in cuprates, consistent with predictions of spin-fluctuation mediated pairing models \cite{LeTacon2011,Huang2017}.

\section{Experimental methods}

LCCO films with a thickness of 10 unit cells (\(\sim 13.2\,\text{nm}\)) and various doping levels were grown on LaSrAlO\(_4\) (LSAO) (001) substrates using molecular beam epitaxy and thoroughly characterized, as previously reported \cite{Kim2021}. Cu $L_3$ resonant inelastic x-ray scattering (RIXS) measurements were carried out using the ERIXS spectrometer at the ID32 beamline of the European Synchrotron Radiation Facility in Grenoble. The scattering angle was fixed at \(149.5^\circ\), and the measurements were performed with an energy resolution of \(\sim 65\,\text{meV}\), using \(\pi\)-polarized incident light ($i.e.,$ polarization parallel to the scattering plane) to enhance the magnetic scattering signal (see Fig.~\ref{fig1}(a)) \cite{Minola2015}. All data were collected at a temperature of 25 K.

\section{Results}

\begin{figure}
	\includegraphics[width=0.49\textwidth]{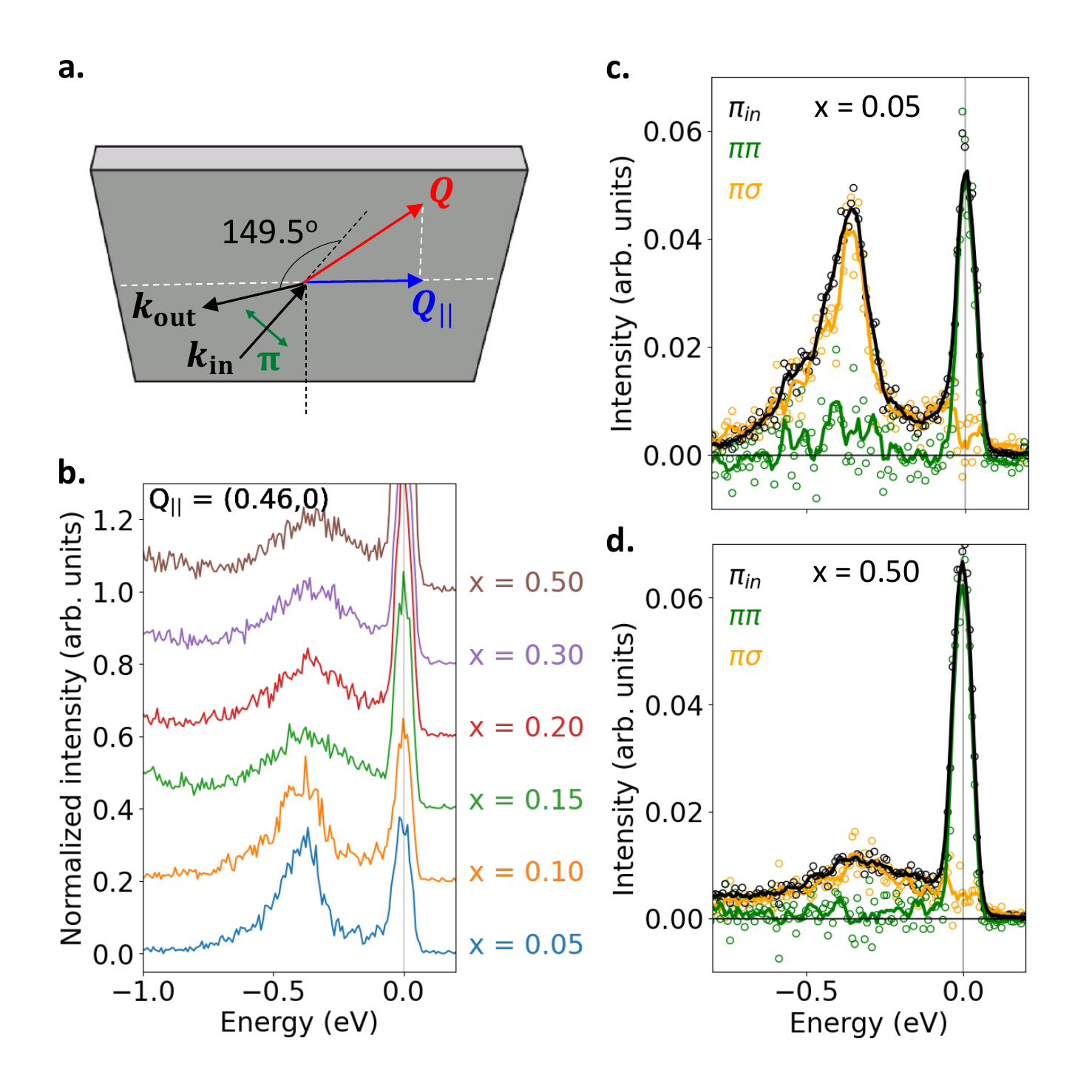}
	\caption{(a) Schematic of the scattering geometry used in the RIXS experiments. (b) RIXS spectra obtained at an in-plane momentum transfer $Q_{||} = (0.46,0)$, for various Ca-doping levels. (c,d) Polarization-resolved RIXS spectra obtained for (c) $x = 0.05$, and (d) $x = 0.50$. The solid lines represent 5-point averaging of the spectra.}
	\label{fig1}
\end{figure}

Figure~\ref{fig1}(a) illustrates the scattering geometry used in the RIXS experiments. As established in previous work~\cite{Minola2015}, a grazing-emission configuration with $\pi$-polarized incident light is optimal for detecting spin-flip excitations via the cross-polarized $\pi\sigma'$ channel. Accordingly, the incident beam was aligned at an angle of approximately \(43^\circ\) relative to the surface normal, resulting in near-grazing emission of the scattered beam and an in-plane momentum transfer of \(Q_{||} = (0.46, 0)~\text{r.l.u.}\). With this geometry, a pronounced mid-infrared peak centered around \(0.35\)–\(0.40~\text{eV}\) is observed across the full doping range from \(x = 0.05\) to \(0.50\) (Fig.~\ref{fig1}(b)). Consistent with Ref.~\cite{Minola2015}, polarization analysis of the scattered beam confirms that the spectral weight in this energy range arises predominantly from the cross-polarized scattering channel, and is therefore dominated by spin-flip excitations (Fig.~\ref{fig1}(c,d)).

\begin{figure}
	\includegraphics[width=0.47\textwidth]{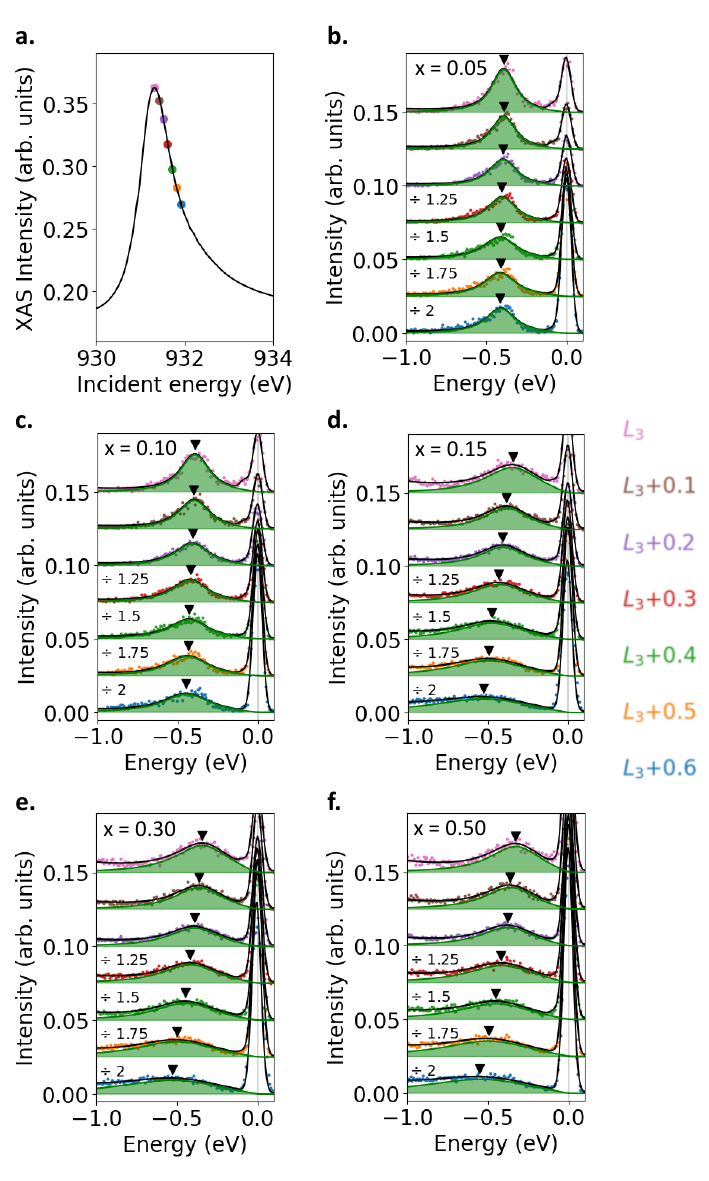}
	\caption{RIXS spectra obtained with selected incident energies detuned from the peak of the Cu-\(L_3\) edge XAS profile (a), for various Ca doping levels: (b) \(x = 0.05\), (c) \(x = 0.10\), (d) \(x = 0.15\), (e) \(x = 0.30\), and (f) \(x = 0.50\). For each doping level, spectra are shown at seven incident energies: one at the maximum of the XAS intensity, and six others shifted upward by 0.1 to 0.6 eV in 0.1 eV steps. Solid black lines in (b--f) represent fits to the data, as described in the text. Green shaded regions indicate the mid-infrared spin-flip component of the fits. Arrows indicate the energy at which the mid-infrared peak reaches maximum intensity. Spectra obtained at energies $L_3 + 0.3$ eV, $L_3 + 0.4$ eV, $L_3 + 0.5$ eV and $L_3 + 0.6$ eV are scaled up by factors of 1.25, 1.5, 1.75 and 2, respectively.} 
	\label{fig2}
\end{figure}

In order to distinguish incoherent spin-flip excitations from collective paramagnons, we examine the evolution of the spin-flip excitation peak as the incident photon energy is detuned from the Cu \(L_3\) resonance (Fig.~\ref{fig2}(a)). Figures~\ref{fig2}(b–f) show the RIXS spectra for \(x = 0.05\), 0.10, 0.15, 0.30, and 0.50 at various detuning levels. Each spectrum is fitted using a combination of components: resolution-limited Gaussians for the elastic line at zero energy loss and the phonon mode near \(\sim 70~\text{meV}\), a Lorentzian background to model the tail of the \(dd\) excitations, and a resolution-convoluted damped harmonic oscillator (DHO) to describe the spin-flip excitation, given by:

\begin{equation}
	I(\omega) = A \cdot \frac{\omega \, \Gamma}{\left( \omega^2 - \omega_0^2 \right)^2 + 4\omega^2 \Gamma^2},
\end{equation}
where \(A\), \(\omega\), \(\omega_0\), and \(\Gamma\) are the amplitude, frequency, undamped frequency, and damping factor, respectively. The resulting fits are shown as solid black lines in Figs.~\ref{fig2}(b–f), with the green shaded regions representing the spin-flip excitation. For the detuned spectra at $x = 0.05$, noticeable deviations from the fit are observed, likely arising from multi-magnon contributions. However, since our primary focus is on the evolution of the peak position with detuning, these deviations do not significantly affect the analysis.

\begin{figure}
	\includegraphics[width=0.4\textwidth]{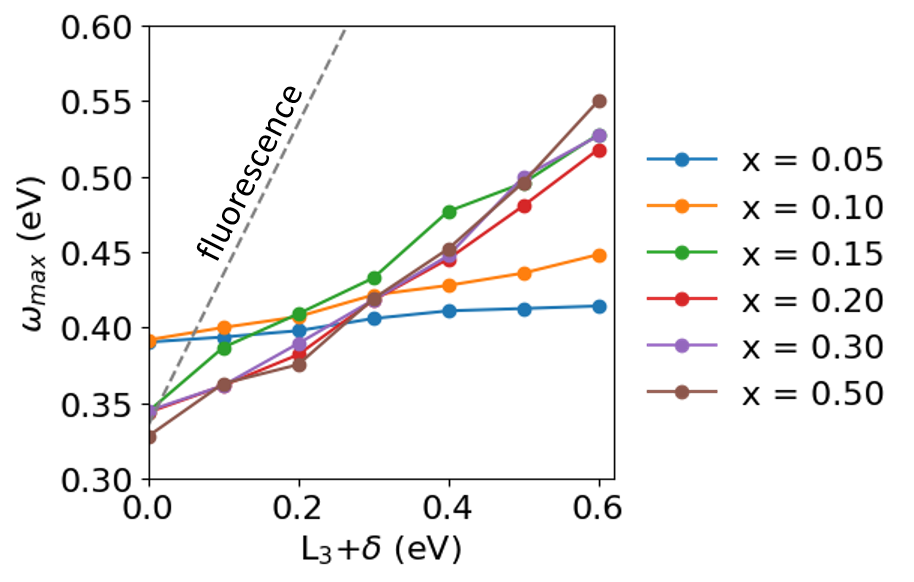}
	\caption{Incident photon energy detuning dependence of the energy corresponding to the maximum intensity of the mid-infrared peak for various Ca doping levels. The dashed line indicates the expected detuning dependence for a fluorescent feature.}
	\label{fig3}
\end{figure}

The extracted energy at which the spin-flip excitation intensity reaches a maximum is shown in Fig.~\ref{fig3}. For \(x = 0.05\) and 0.10, a nearly constant peak energy across all detuning levels fits the data well—indicative of Raman-like behavior—which clearly suggests a collective spin-excitation origin. In contrast, for dopings in the range \(x = 0.15\)–0.50, the peak position shifts noticeably with increasing detuning. However, the detuning dependence of \(\omega_{\text{max}}\) remains clearly distinct from that expected for purely fluorescent behavior, as indicated by the dashed line in Fig.~\ref{fig3}.

\begin{figure}
	\includegraphics[width=0.45\textwidth]{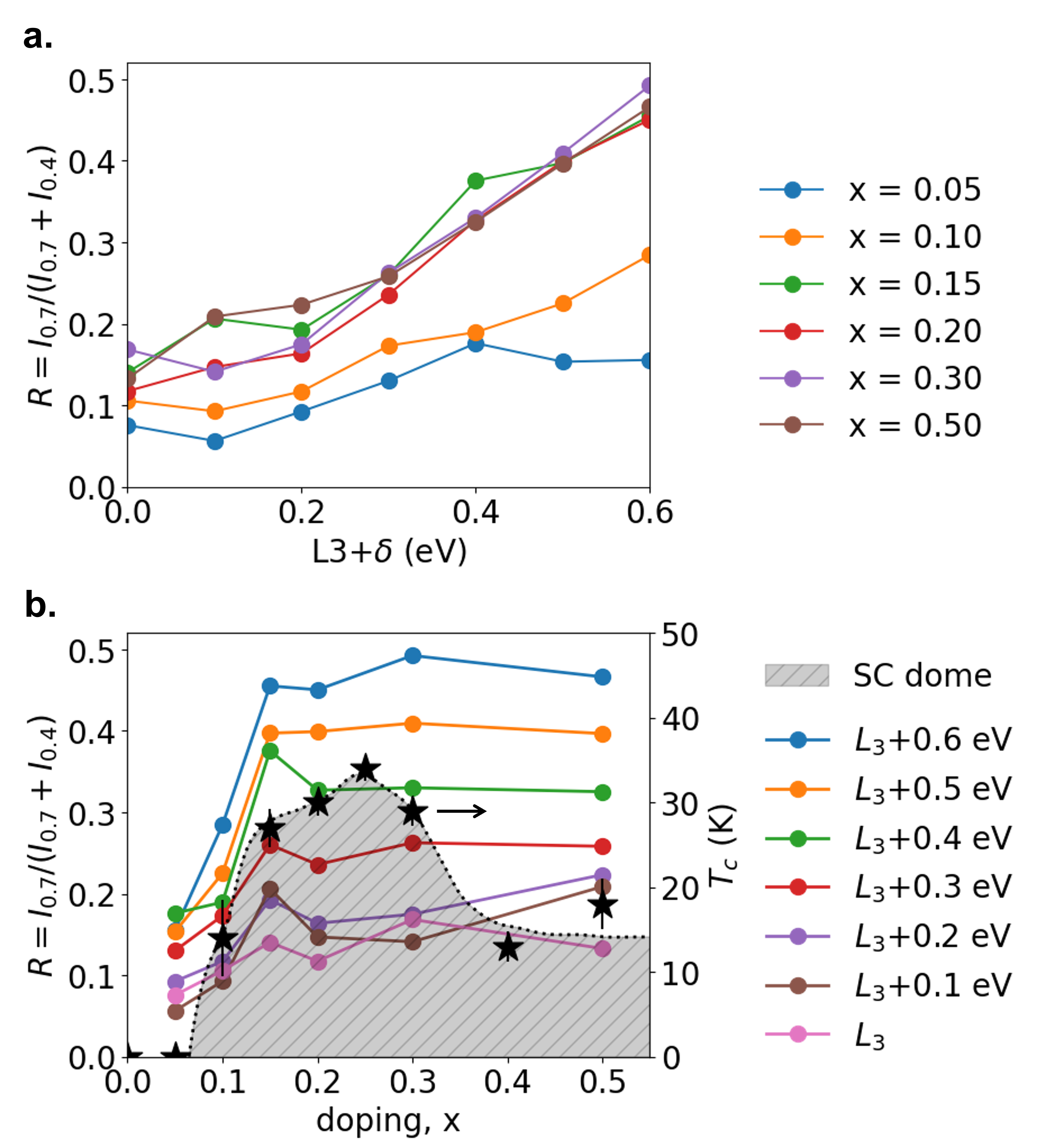}
	\caption{Ratio of averaged RIXS intensities within a 25 meV window centered at energy transfers of 0.7 eV and 0.4 eV, shown in (a) as a function of detuning for various Ca doping levels, and in (b) as a function of Ca doping for various detuning levels. The superconducting dome is included in (b) for comparison.}
	\label{fig4}
\end{figure}

\begin{figure*}
	\includegraphics[width=\textwidth]{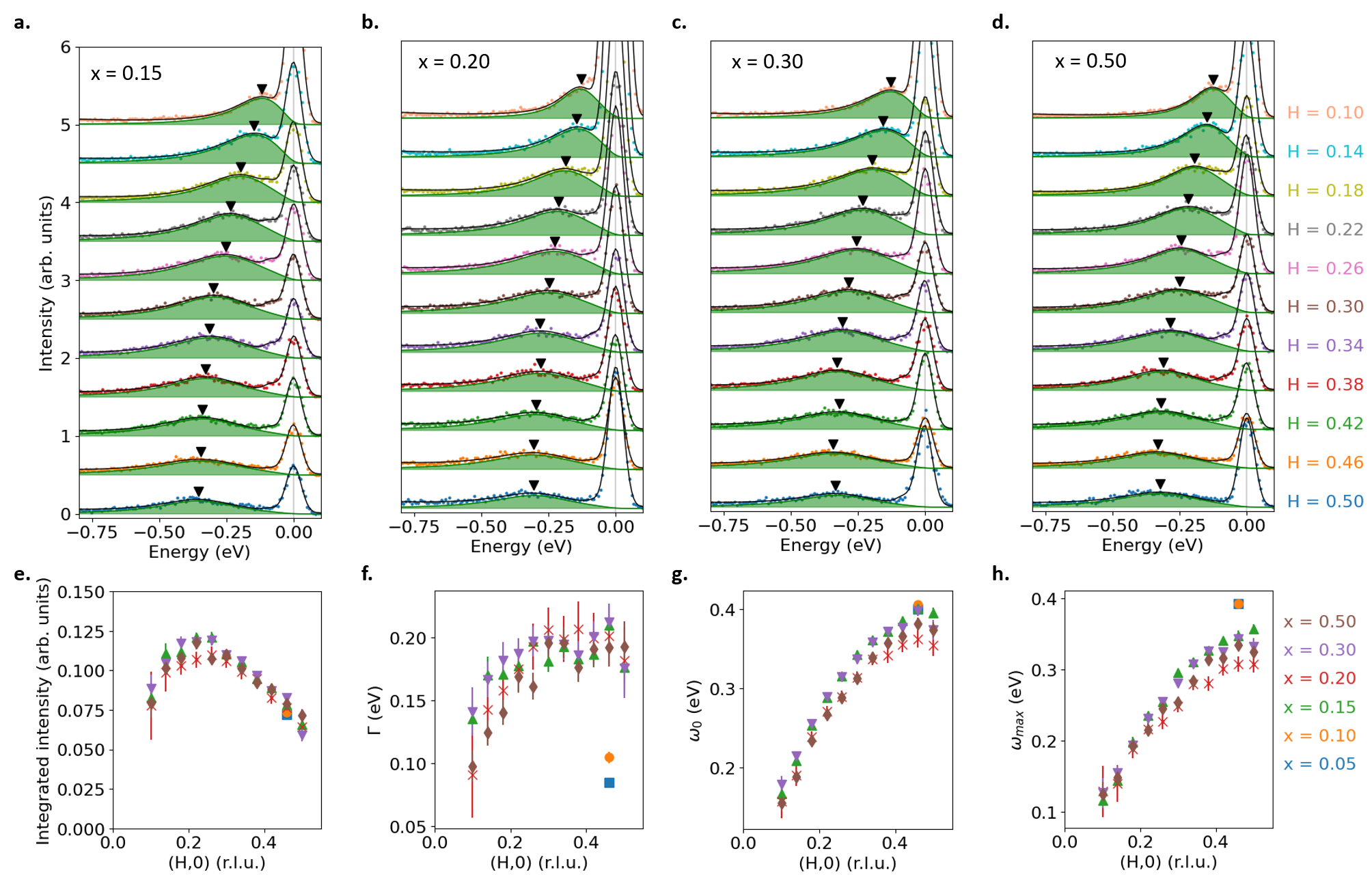}
	\caption{Momentum dependence of the RIXS spectra along \( Q_{||} = (H,0) \) for various Ca doping levels: (a) \( x = 0.15 \), (b) \( x = 0.20 \), (c) \( x = 0.30 \), and (d) \( x = 0.50 \). The solid black lines represent fits to the spectra as detailed in the text. The green shaded regions correspond to the spin-flip excitation component of the fits. Black triangles indicate the energies where the spin-flip excitation intensity is maximal. Panels (e–h) show parameters extracted from these fits (including data from Fig.~\ref{fig2}(b,c)): (e) integrated intensity, (f) damping \(\Gamma\), (g) undamped frequency \(\omega_0\), and (h) frequency at which the spin-flip excitation intensity is maximal \(\omega_{\text{max}}\).}
	\label{fig5}
\end{figure*}

Prior numerical calculations based on the two-dimensional single-band Hubbard model have shown that for the specific grazing-emission geometry used in our experiment, RIXS spectra in the hole-doped system comprise two components with distinct detuning dependencies: a paramagnon feature with a peak energy that remains independent of detuning (Raman-like behavior), and a particle-hole continuum feature that shifts to higher energy with increasing detuning (fluorescence-like behavior)~\cite{Minola2017}. Experimentally, this picture has been validated across several families of cuprates, where the mid-infrared spin-flip excitation feature exhibits partial fluorescence behavior around optimal doping (\(p \sim 0.16\)), due to increased contributions from particle-hole continuum excitations \cite{Minola2017}. Here, we investigate whether this framework applies to the present system. Following previous work~\cite{Minola2017}, we employ the ratio \(R\) of the intensity at the spin-excitation peak (\(\sim 0.4\)~eV) to that on the high-energy tail (0.7~eV) as a proxy for the integrity of the collective spin excitation. This ratio remains small when the spectral weight is concentrated in a well-defined collective mode, but increases as the excitation loses spectral weight to the incoherent continuum.

Figure~\ref{fig4}(a,b) shows the dependence of the ratio \( R \) on detuning and Ca doping, respectively. For \( x = 0.05 \) and 0.10, the dominant spectral weight originates from a collective spin excitation, as indicated by the consistently small values of \( R \) across all detuning levels. In contrast, a pronounced crossover occurs around \( x = 0.15 \), where \( R \) increases significantly with detuning. This behavior implies a suppression of collective mode spectral weight in favor of incoherent spin-flip excitations within the particle-hole continuum. For \( x \geq 0.15 \), the spectra evolve only weakly with further doping. Thus, while a clear crossover is observed near \( x = 0.15 \), where RIXS intensities increasingly reflect incoherent spin-flip excitations, a collective spin-excitation component with a Raman-like detuning dependence appears to persist up to Ca doping levels as high as \( x = 0.50 \). This behavior contrasts with previous theoretical predictions that attribute the entire mid-infrared spectral weight to incoherent particle-hole excitations \cite{Benjamin2014,Nagy2016}. The persistence of a residual paramagnon component prevents the RIXS intensities from exhibiting fully fluorescent behavior (i.e., \( R = 1 \)) with detuning. 

It is important to note that the collective mode and incoherent spin-flip excitations cannot be cleanly separated in the RIXS spectra for \( x = 0.15\text{--}0.50 \), as for these dopings, the mid-infrared spin-flip excitation peak can be modeled using a single resolution-convoluted DHO function. Nevertheless, we have measured the dispersion of this feature for \( x = 0.15\text{--}0.50 \), as presented in Fig.~\ref{fig5}(a--d). The spectra are fit using the same model as for the detuning data in Fig.~\ref{fig2}; \emph{i.e.}, a combination of resolution-limited Gaussians for the elastic line at zero energy loss and the phonon mode near \(\sim 70~\text{meV}\), a Lorentzian background to account for the tail of the \(dd\) excitations, and a resolution-convoluted DHO to describe the spin-flip excitation. The integrated intensity, damping factor (\(\Gamma\)), undamped frequency (\(\omega_0\)), and the frequency corresponding to the peak intensity of the spin-flip component (\(\omega_{\text{max}}\)) obtained from the fits are displayed in Fig.~\ref{fig5}(e--h). The fitted parameters for \( x = 0.05 \) and 0.10, obtained at \( Q_{||} = (0.46, 0) \) from the Cu \( L_3 \)-edge resonance data in Fig.~\ref{fig2}(b,c), are also included in Fig.~\ref{fig5}(e--h) for comparison. The results are broadly consistent with prior reports of paramagnons in overdoped cuprates—namely, an overall broadening of the spin-flip excitation peak is observed with increasing doping, accompanied by no significant change in either the integrated spectral weight or the characteristic energy across the doping range~\cite{LeTacon2011,LeTacon2013,Dean2013,Peng2018,Zhang2022}. However, the presence of overlapping incoherent excitations precludes fully unambiguous conclusions regarding the doping dependence of the paramagnon spectral weight or energy.

\section{Discussion and Conclusions}

Our RIXS measurements on LCCO reveal that collective paramagnon excitations persist up to the highly overdoped regime (\( x = 0.50 \)), with a crossover occurring at \( x = 0.15 \), where a loss of paramagnon spectral weight to incoherent spin-flip excitations is observed. Despite the extended superconducting dome in LCCO, the paramagnon behavior is strikingly similar to that observed in other hole-doped cuprate families such as (Bi,Pb)$_2$(Sr,La)$_2$CuO$_{6+\delta}$ (Bi2201), Tl$_2$Ba$_2$CuO$_{6+\delta}$ (Tl2201), and YBCO, which exhibit more conventional superconducting domes~\cite{Minola2015,Minola2017}. 

Our findings are broadly consistent with determinant quantum Monte Carlo calculations on the single-band Hubbard model, which predict that both the energy and spectral weight of high-energy spin excitations near the antiferromagnetic zone boundary remain largely unchanged with hole doping~\cite{Jia2014,Huang2017}. Moreover, phenomenological models \cite{LeTacon2011} and numerical calculations \cite{Huang2017} indicate that the suppression of superconductivity in the overdoped regime is more closely associated with a depletion of low-energy spin excitations near the antiferromagnetic zone center. This loss of low-energy spectral weight has clearly been observed in neutron scattering experiments~\cite{Wakimoto2004,Wakimoto2007,Lipscombe2007}, providing experimental support for the theoretical picture. Our RIXS measurements align well with this scenario: while high-energy magnetic excitations persist across the doping range, they show no clear correlation with the extended superconducting dome in LCCO, reinforcing the idea that these high-energy excitations probed by RIXS play only a limited role in the pairing mechanism. Note that RIXS measurements are limited in accessible momentum transfer to approximately twice the photon momentum, which is about $0.943$~\AA$^{-1}$ at the Cu $L_3$ edge ($\sim$931~eV). This is well below the momentum transfer corresponding to the antiferromagnetic zone center ($\sim$ $1.18$~\AA$^{-1}$), and therefore RIXS cannot access the low-energy excitations observed by neutron scattering.

Our results underscore the challenges involved in interpreting the doping dependence of paramagnons, particularly in the intermediate and overdoped regime (\(x = 0.15\) and above), with RIXS. Due to the strong contribution from incoherent spin-flip excitations for \(x \geq 0.15\), any straightforward comparison of peak energies across the full doping range risks oversimplifying the underlying physics. 
Future methodological advances—particularly follow-up calculations and theoretical modeling informed by the present and prior experimental datasets \cite{Minola2015,Minola2017}—may enable a clearer separation between the incoherent and collective paramagnon components of the spin-flip excitation peak.

Prior studies~\cite{Minola2015,Minola2017} and our current results suggest that the transfer of spectral weight from collective spin excitations—characteristic of the undoped Mott insulator—to particle-hole continuum excitations, typical of an uncorrelated Fermi liquid, occurs near \( x \approx 0.15 \). As noted in previous work \cite{Minola2017}, this doping level closely coincides with the hole-doping where transport and angle-resolved photoemission spectroscopy (ARPES) measurements indicate a major reconstruction of the Fermi surface, with the hole-carrier density gradually increasing from \( p \) to \( 1 + p \) \cite{Fournier2010,Badoux2016,Putzke2021,Nicholls2025}. The continuum excitations observed in the present study, as well as in earlier investigations of other cuprates \cite{Minola2015,Minola2017}, may be associated with the emergence of these additional carriers. Recent theoretical work has proposed that the spin-flip excitation peak observed in RIXS in doped cuprates could be related to a spinon continuum, potentially explaining the missing Fermi surface volume in the pseudogap phase \cite{Bonetti2024,Bonetti2025}. However, the predicted continuum is significantly broader than the experimentally observed spin-flip peak. Moreover, the peak remains well-defined deep into the overdoped regime in LCCO and other cuprates such as YBCO, Bi2201, and Tl2201~\cite{Minola2015,Minola2017}. These observations argue against a spinon continuum as the origin of the spin-flip peak.

In conclusion, our work demonstrates that the doping dependence of the integrity of high-energy spin excitations in LCCO is similar to that in other cuprate families and interestingly is largely unaffected by the extension of the superconducting dome in LCCO into the extremely overdoped regime. These findings lend support to the view that high-energy paramagnons probed by RIXS are not the primary driving mechanism of superconducting pairing in the cuprates.

\section{Acknowledgments}
We thank Pietro Maria Bonetti, Alex Nikolaenko, and Subir Sachdev for helpful discussions. The RIXS data were collected at the beam line ID32 of the European Synchrotron Radiation Facility (ESRF) in Grenoble, France (proposal \# HC-5428). S. H. is supported by the Alexander von Humboldt Foundation.

\bibliographystyle{apsrev4-2}
\bibliography{LCCO_PM.bib}

\end{document}